\documentclass[aps,pre,reprint,superscriptaddress]{revtex4-1}
\usepackage{amsmath}
\usepackage[latin1]{inputenc}
\usepackage{graphicx}

\usepackage{epsfig}

\usepackage{color}

\begin{document}

\title{Percolation of functionalized colloids on patterned substrates}

\begin{abstract}
	We study the percolation properties for a system of functionalized
	colloids on patterned substrates via Monte Carlo simulations. The
	colloidal particles are modeled as hard disks with three
	equally-distributed attractive patches on their perimeter. We describe the
	patterns on the substrate as circular potential wells of radius $R_p$
	arranged in a regular square or hexagonal lattice. We find a
	nonmonotonic behavior of the percolation threshold (packing fraction)
	as a function of $R_p$.  For attractive wells, the 
	percolation threshold is higher than the one for clean
	(non-patterned) substrates if the circular wells are non-overlapping
	and can only be lower if the wells overlap. For repulsive wells we
	find the opposite behavior. In addition, at high packing fractions the formation
	of both structural and bond defects suppress percolation. As a result, the
	percolation diagram is reentrant with the non-percolated state occurring at very low
	and intermediate densities.
\end{abstract}

\author{Lucas L. Treffenst{\"a}dt}
\affiliation{Theoretische Physik II, Physikalisches Institut,
  Universit{\"a}t Bayreuth, D-95440 Bayreuth, Germany}
\author{Nuno A.M. Ara\'ujo}
\affiliation{
Departamento  de  F{\'i}sica,  Faculdade  de  Ci{\^e}ncias,  Universidade  de  Lisboa,
P-1749-016  Lisboa,  Portugal,  and  Centro  de  F{\'i}sica  Te\'orica  e  Computacional,
Universidade  de  Lisboa,  P-1749-016  Lisboa,  Portugal}
\author{Daniel de las Heras}
\email{www.danieldelasheras.com}
\affiliation{Theoretische Physik II, Physikalisches Institut,
  Universit{\"a}t Bayreuth, D-95440 Bayreuth, Germany}

\date{\today}

\maketitle

\section{Introduction}

Substrates patterned with space-dependent physico-chemical properties have a
strong influence on the equilibrium and dynamical properties of soft materials
deposited or adsorbed on them. Examples of physical processes affected by the
presence of patterns include
wetting~\cite{PhysRevLett.80.1920,PhysRevE.60.6919}, crystal
nucleation~\cite{aizenberg1999control,1367-2630-15-7-073013}, phase
separation~\cite{boltau1998surface},
freezing~\cite{PhysRevLett.85.3668,doi:10.1063/1.3383239},
adsorption~\cite{Araujo2008,Araujo2017},
sedimentation~\cite{PhysRevE.79.011403}, and colloidal
transport~\cite{PhysRevLett.112.048302,paper31,paper33}.

Patterned substrates have been considered as a potential route to control the
self-assembly of colloidal
particles~\cite{doi:10.1021/la010682j,doi:10.1021/la000277c,doi:10.1021/cr400081d,B311283G,Cadilhe2007}.
For example, even simple one-dimensional periodic patterns can induce the
formation of chains, regular lattices~\cite{doi:10.1063/1.1391234} or even more
complex structures~\cite{PhysRevE.65.041602}. In parallel, another potential
route that has been considered to colloidal self-assembly is the use of particles with
functionalized patches (patchy particles), yielding directional interactions
and limited valence~\cite{in1,in2}. Patchy colloids are ideal building blocks to obtain
e.g. empty liquids~\cite{emptyscio,paper11,paper12,emptyexp}, colloidal
micelles~\cite{kraft2012surface}, quasicrystals~\cite{C6SM01838F}, and complex
lattices~\cite{chen2011directed,mcmcmc}.

The first studies of equilibrium and non-equilibrium properties of
functionalized colloidal particles on substrates reveal that, even on clean
(non-patterned) substrates, the critical behavior and dynamics depend strongly on
the particle-particle and particle-substrate interaction and number of
patches, as reviewed in Ref.~\cite{Dias2017}. Here, we focus on the
equilibrium properties of functionalized colloids in the presence of a patterned substrate. We consider
simple patterns consisting of circular wells regularly distributed in a square
or hexagonal lattice arrangement. We focus on the percolation properties and
show that the critical packing fraction for percolation depends strongly on
the nature of the particle-well interaction (attractive or repulsive) and on the
radius of the wells, $R_p$. In particular, the percolation transition can
either be delayed or anticipated, when compared to the critical packing
fraction on a clean substrate.  For a wide range of model parameters, we find
a reentrant percolation transition driven by the formation of both structural and
bond defects.

\section{Model and methods}

\begin{figure}
\includegraphics[width=1.0\columnwidth]{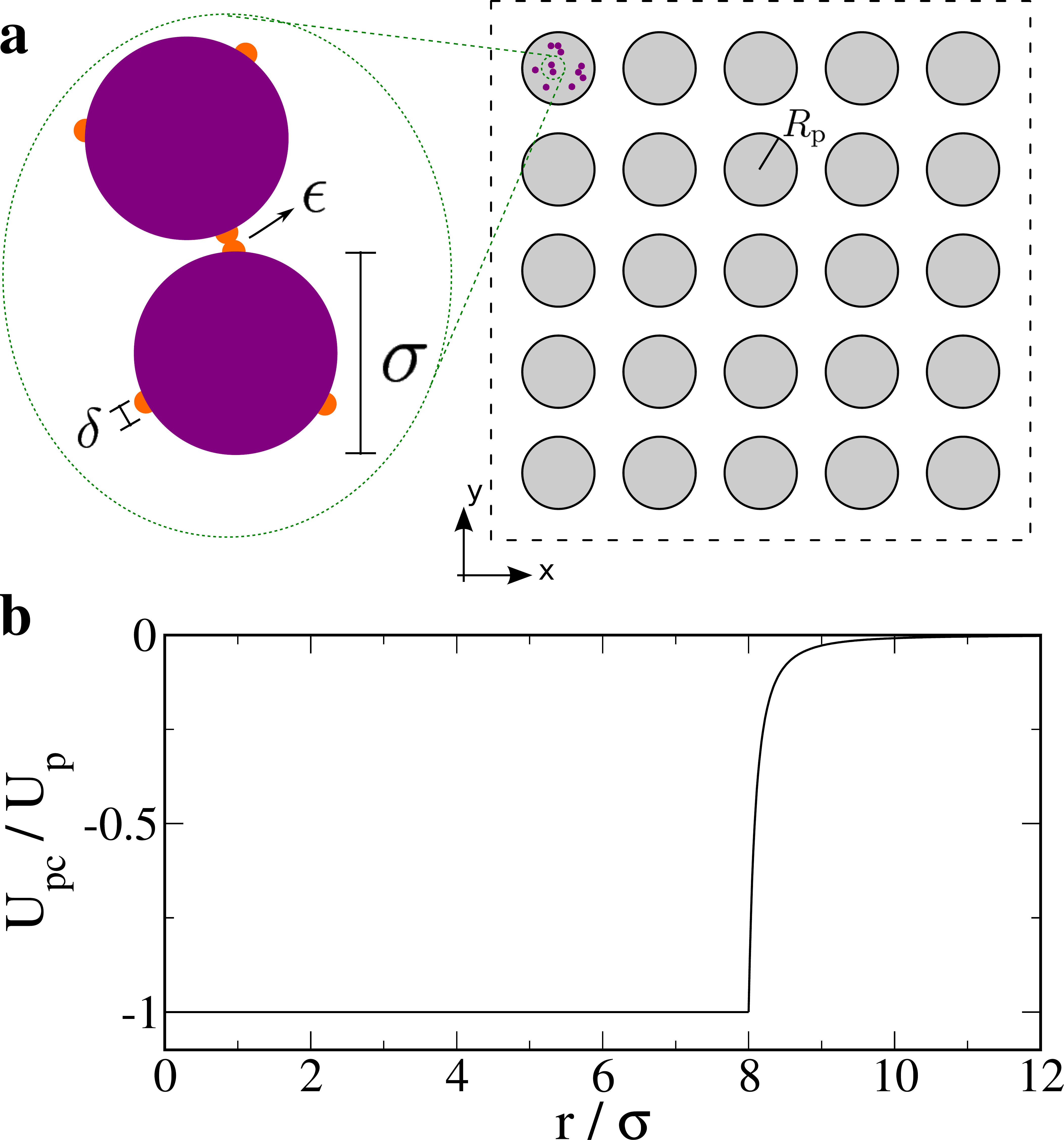}
\caption{(a) Schematic of the model: a substrate patterned with a square lattice
	of potential wells (grey circles of radius $R_p$). The interaction
	potential between the wells and the colloidal particles (disks) is
	either attractive or repulsive. In the case of attractive wells, the
	particles are more likely to be found inside the wells, as illustrated in
	the figure. The inset shows a close view of the particle model.
	Colloidal particles are described as hard-core disks of diameter $\sigma$
	decorated with three patches (disks of diameter $\delta$) equally
	distributed on their surface. Whenever two patches overlap the
	potential energy decreases by $\epsilon$. (b) Plot of the radial
	dependence of the particle-well potential as a function of the distance
	$r$ between the centers of the particle and the well.
	\label{fig1}}
\end{figure}

As summarized in Fig.~\ref{fig1}(a), the functionalized colloidal particles are
modeled as hard disks (core) of diameter $\sigma=1$ with three patches equally
distributed on their surface (perimeter). The patches are described as smaller
disks of diameter $\delta/\sigma=0.5(\sqrt{5-2\sqrt{3}}-1)\approx 0.120$.  The
core-core interaction is hard and the patch-patch interaction is
such that the potential energy decreases by $\epsilon$ when two patches
(partially) overlap, independently of the overlapping area. For the considered
size and arrangement of the patches: (i) only bonds involving two patches are
possible and (ii) a pair of particles can share at most one single bond~\cite{sciortino2d}. 

The substrate is patterned with circular wells of radius $R_p$ with their
geometric centers distributed spatially either in a square or hexagonal lattice
arrangement. Particles are distributed on the substrate forming a monolayer.
The particle-substrate interaction potential, $U_{\text{pc}}(r)$, is radial, solely
depending on the distance $r$ between the centers of the particle and the well.
$U_{\text{pc}}(r)$ is constant for a core $r<R_p$ and goes to zero with the inverse
square distance for $r>R_p$, see Fig.~\ref{fig1}(b). For numerical simplicity, we introduce
a cut-off distance, $R_c$, which is fixed at half of the lateral length of the simulation box. The
potential $U_{\text{pc}}(r)$ is slightly shifted such that it is continuous at
$R_c$, so
\begin{equation}
	\frac{U_{\text{pc}}(r)}{U_p}= \left\{
	\begin{array}{ll}
		-1 & \text{if } r < R_p\\
		0 & \text{if } r > R_c\\
		-\frac{a^2}{\left(r+a-R_p\right)^2}+\frac{a^2}{\left(R_c+a-R_p\right)^2}\frac{r-R_p}{R_c-R_p}
		& \text{otherwise.} \\
	\end{array} \right.
	\label{eqn:u1}
\end{equation}
The parameter $a$ controls the softness of the potential and we fix it at
$a/\sigma=0.2$. The depth of the potential well is given by $U_{p}$.  We set
$U_{p} / \epsilon =-4$ ($U_{p} / \epsilon =4$) for the case of repulsive
(attractive) potential wells. We consider $25$ and $36$ wells for square and
hexagonal lattices, respectively. We vary the size of wells from very small
($R_p\approx\sigma$) non-overlapping wells to very large ($R_p\gg\sigma$) fully
overlapping wells. For the case of overlapping wells, the potential inside the
overlapping region is always set to $U_p$.
For both patterns, the lattice
constant is $20\sigma$. Thus, the lateral length of the substrate is
$L\approx100\sigma$. We use periodic boundary conditions along both
directions.

We performed canonical Monte Carlo simulations to study the percolation
properties. The control parameter is the packing fraction, $\eta=Nv_0/A$,
where $N$ is the number of particles, $v_0=\pi(\sigma/2)^2$ is the volume
(area) of the hard cores, and $A$ is the total area of the substrate.
Here, we consider values of $N$ in the range $\approx[10^3,10^4]$, and thus
$\eta\approx[0.1,0.7]$. At each Monte Carlo sweep (MCS), we sequentially perform an
attempt to move and rotate every particle. The displacements and the rotation angles
are randomly generated from a uniform distribution. 
At the beginning of each simulation we
estimate the maximum displacement and maximum rotation angle that 
each particle is allow to perform in one move. Both parameters are adjusted such that
approximately $50\%$ of all particles moves are accepted during the simulation.

To generate the initial configuration, we distribute the particles without
overlapping and then run $10^4$ MCS at a very high temperature ($k_BT/\epsilon=10^2$) 
such that the positions and the orientations of all particles are randomized. 
We then equilibrate the system running $5\times10^6-10^7$ MCS (depending on the model
parameters) and accumulate data over $10^6-10^7$ additional MCS.

\section{Results} 
In bulk, a two-dimensional system of functionalized particles with three
patches undergoes a first order vapor-liquid transition by increasing the
density at sufficiently low temperatures~\cite{sciortino2d}. The transition
line ends at a critical point for high temperatures. Above the critical
temperature, a continuous percolation transition is still observed by increasing
the particle density.

\begin{figure}
\includegraphics[width=0.9\columnwidth]{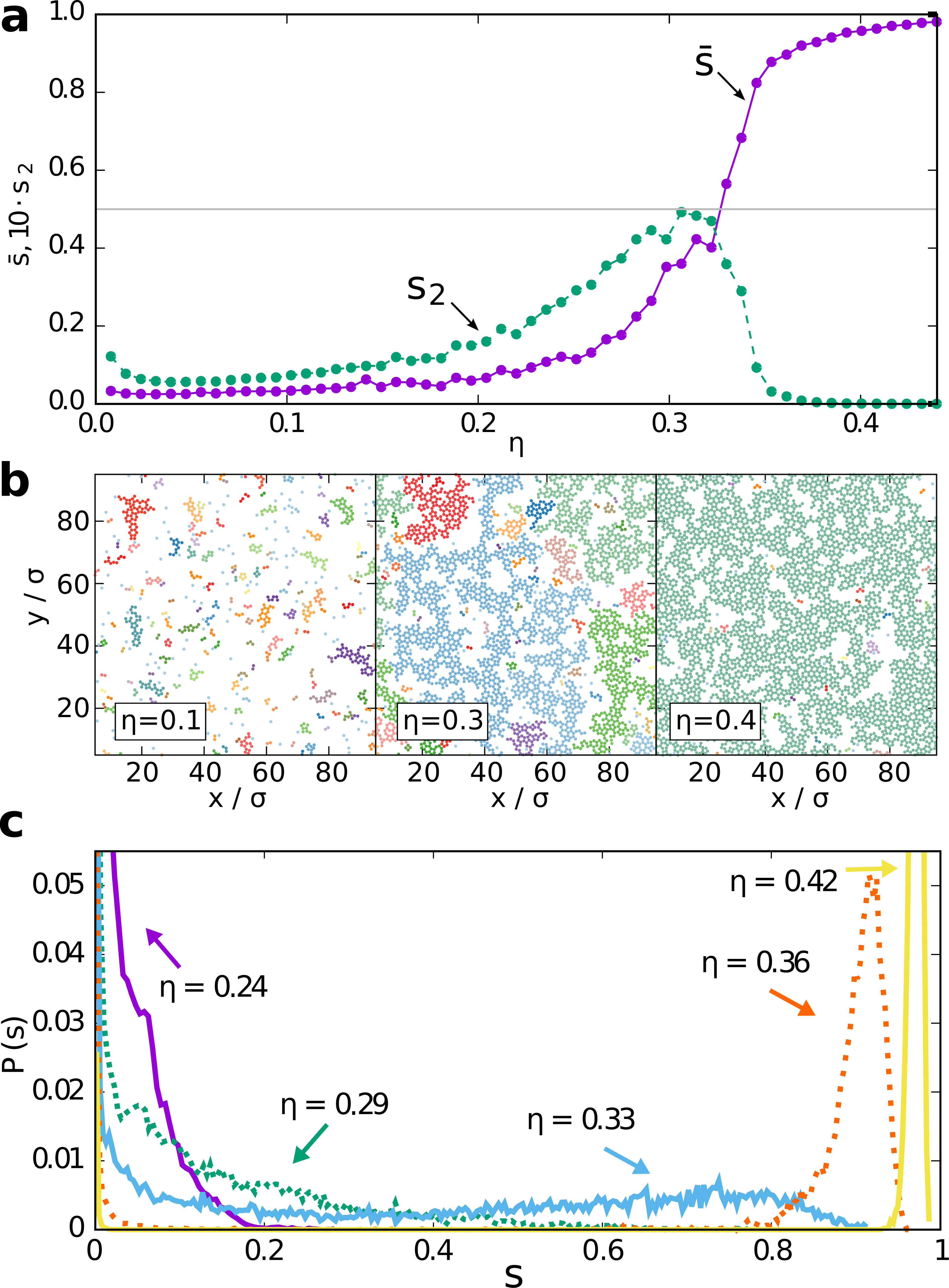}
\caption{Percolation on a clean substrate. (a) Fraction of particles in the
	largest cluster $\bar s$ and second moment of the cluster-size
	distribution $s_2$ (multiplied by a factor $10$ for visualization
	purposes) as a function of the packing fraction. (b) Snapshots for
	three packing fractions: below $\eta=0.1$ (left), close to $\eta=0.3$
	(middle), and above $\eta=0.4$ (right) the percolation transition.
	Each cluster is colored differently. (c) Cluster size distribution
	$P(s)$, as a function of the cluster size $s$, for different values of
	the packing fraction, as indicated.  \label{fig2}}
\end{figure}

Here, to focus on the role of the pattern, we fix the temperature
at $k_{B}T/\epsilon=0.15$, which is well above the vapor-liquid critical point
of the bulk phase diagram~\cite{sciortino2d}. We first analyze the percolation
transition on a clean substrate, i.e., a substrate with no potential wells.
To characterize the percolation transition, we measure the fraction $\bar s$
of particles belonging to the largest cluster of connected particles. For
simplicity, we estimate the percolation threshold as the value of the packing
fraction $\eta=\eta_0$ at which $\bar s=1/2$. As shown in Fig.~\ref{fig2}(a),
this roughly coincides with the position of the peak in the second moment of the
cluster-size distribution. We
estimate $\eta_0\approx0.33$ for a clean substrate.  Snapshots for packing
fractions below, close, and above the percolation threshold are shown in
Fig.~\ref{fig2}(b). Figure~\ref{fig2}(c) shows the cluster-size distribution
$P(s)$, where $s$ is the fraction of particles in the cluster.  At low
packing fractions the particles are aggregated in small (finite) clusters,
while above the percolation threshold, there is an infinite cluster that spans
the entire substrate. 

\subsection{Percolation on a patterned substrate: square lattice}
\begin{figure}
\includegraphics[width=1.0\columnwidth]{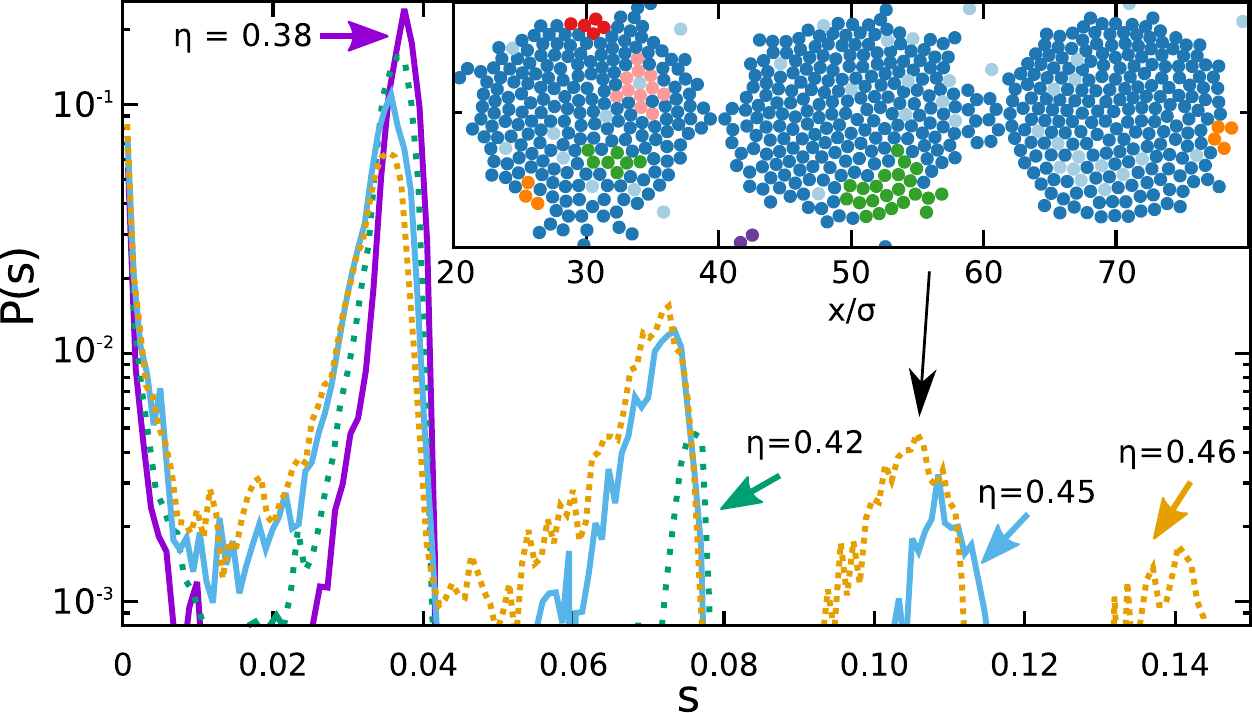}
	\caption{Semi-logarithmic plot of the cluster-size distribution $P(s)$
	as a function of the cluster size $s$ for different values of the
	packing fraction, as indicated. The substrate is patterned with a
	square lattice of circular potential wells of radius $R_p/\sigma=8$.
	The inset is a partial snapshot of the simulation box ($\eta=0.45$)
	showing a cluster of colloidal particles that extends over three
	potential wells. Different particle colors correspond to different
	clusters.~\label{fig3}}
\end{figure}

Let us first consider a pattern consisting of attractive circular wells
arranged in a square lattice. Figure~\ref{fig3} shows the cluster-size
distribution, $P(s)$, for different packing fractions, when the radius of the
wells is $R_p/\sigma=8$. Clearly, $P(s)$ is significantly different from the
one obtained on a clean substrate, see Fig.~\ref{fig2}(c). $P(s)$ is
characterized by a sequence of well defined peaks for cluster of sizes $s\approx ns_0$.
Here, $s_0$ is the characteristic size of a cluster of radius $R_p$ and
$n=1,2,3,\dots$. As shown in the inset of Fig.~\ref{fig3}, the colloidal particles
tend to accumulate inside the attractive wells, what promotes the formation of
compact clusters. As the average packing fraction increases, not all particles
can fit within the potential well and thus they occupy the interstitials
between the wells.  Nevertheless, since the particle-well interaction potential 
decays with the distance to the center, the particles tend to accumulate close to
the perimeter defined by $R_p$. For repulsive wells, the colloidal particles
tend first to occupy the interstitials and only for high average packing
fractions occupy the core of the wells.

\begin{figure*}
\includegraphics[width=0.8\textwidth]{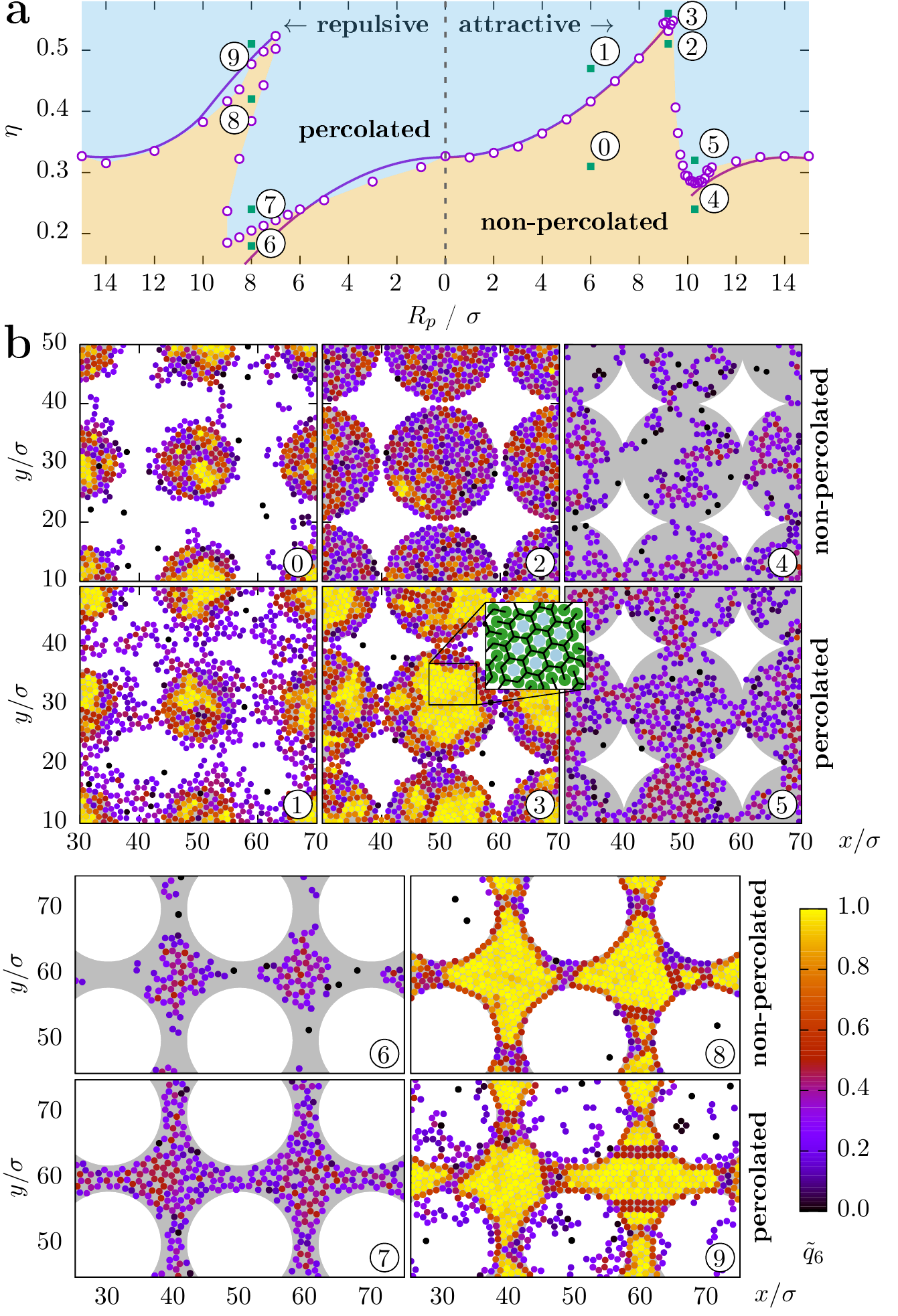}
\caption{Square lattice. (a) Percolation phase diagram in the plane of packing
	radius of the potential well core $R_p$ and packing fraction $\eta$. The light
	brown (light blue) area is the non-percolated (percolated) region, as
	indicated. On the left (right) of the vertical dashed line, we show
	the results for repulsive (attractive) wells. The wells on the
	substrate form a square lattice. The numbered green squares are
	selected state points for which a snapshot is shown in panel (b). The
	empty circles are the percolation values estimated numerically using
        Monte Carlo simulations. The
	solid line is the theoretical prediction (see text for details). (b)
	Selected snapshots of the central region of the substrate.  The
	packing fractions and radii of the well of each configuration is
	indicated in panel (a) (green circles). The top (bottom) group of
	snapshots correspond to attractive (repulsive) wells. The grey areas
	indicate the energetically favorable regions for the colloidal cores.
	The colloids are colored according to the value of their order parameter
	$\tilde q_6$. The inset in the snapshot $3$ is a close view showing
	the underlying bonding lattice: a honeycomb lattice with an isolated
	particle in each unit cell.\label{fig4}}
\end{figure*}

Figure~\ref{fig4} summarizes our findings for the percolation properties on a
substrate with potential wells organized in a square lattice.
Figure~\ref{fig4}(a) is the two-parameter percolation diagram (packing fraction
$\eta$ vs. radius of the well $R_p$), for attractive (right-hand side) and
repulsive (left-hand side) wells. The system exhibits a very rich behavior
that we discuss in detail below.

{\bf Attractive wells.} We start describing the limit of small non-overlapping
attractive wells. In this case, the colloidal particles tend to accumulate
inside the wells. Only when the density inside the wells is significantly
high, particles occupy the space in between wells, eventually forming bridges
between clusters in different wells, what leads to global percolation.
Therefore, the packing fraction at the percolation threshold, $\eta_p$, is
larger than the one for clean substrates $\eta_0$.

To characterize the structure of the clusters, we calculate the orientational
order parameter
\begin{equation}
\tilde q_6^{(i)}=\left|\frac16\sum_{j=1}^{n_i}\exp(i6\theta_j^{(i)})\right| \
	\ ,
\end{equation}
where, $\theta_j^{(i)}$ is the angle between an arbitrary axis of reference
and the vector joining the geometrical centers of particles $i$ and $j$. The
sum runs over all particles at a distance $r/\sigma<1.2$ from particle $i$.
This order parameter is one when particle $i$ is surrounded by six neighbors in a
hexagonal configuration. 

Snapshots below and above the percolation transition in the case of small non-overlapping
wells are shown in Fig.~\ref{fig4}(b), state points $0$ and $1$, respectively.
(See also Fig.~\ref{fig4}(a) to locate the state points in the percolation diagram.)
The particles are colored according to their order parameter $\tilde q_6$.

At the percolation transition, the particle density inside the wells is much
higher than the percolation density on a clean substrate. Inside the wells,
particles are organized in a six-fold symmetry, corresponding to a high
value of the order parameter $\tilde q_6$. However, not all patches are bonded
due to geometrical constraints. At very high packing fractions, particles in
the largest, percolated cluster are fully bonded, forming a honeycomb-like lattice with an
isolated (not bonded) particle in the middle, as shown in the inset of
Fig.~\ref{fig4}(b), for the state point $3$. In the interstitials of the wells,
the particles form an open network (note the low value of $\tilde q_6$),
similar to what is observed for clean substrates close to the percolation
threshold. 

Based on the observations described above, we can estimate the packing
fraction at the percolation threshold. For that, we hypothesize that (i) the
packing fraction inside the wells is $\eta_1>\eta_0$; (ii) the packing
fraction in the interstitials is the same as the one for a clean substrate,
$\eta_0$. Hence, the percolation threshold, for non-overlapping attractive
wells is
\begin{equation}\label{eq::eta}
\eta_{\text{a,no}}(R_p)=A_{\text{no}}\eta_1+(1-A_{\text{no}})\eta_0,
\end{equation}
where $A_{\text{no}}$ is the fraction of the area of the substrate covered by
the core of the wells (within $R_p$). Thus, for non-overlapping wells,
$A_{\text{no}}=n_p\pi R_p^2/A$, with $n_p$ the number of wells and $A$ the area
of the substrate. As shown in Fig.~\ref{fig4}(a) (solid lines),
Eq.~(\ref{eq::eta}) with $\eta_1=0.65$ is a good estimator for the percolation
line. In the limit  $A_{\text{no}}\rightarrow0$ (very-small wells),  the
packing fraction at the percolation threshold converges to the one on a clean
substrate, i.e.,
\begin{equation}
	\lim_{R_p\rightarrow0}\eta_p(R_p)=\eta_0.
\end{equation}
The packing fraction at the percolation threshold increases monotonically with
$R_p$ until $R_p/\sigma\approx9$. For this value of $R_p$,
$\eta_p\approx0.54$, which is about $1.6$ times higher than the one on a clean
substrate. Above that value of $R_p$, an abrupt decay of $\eta_p$ is observed.
From $R_p/\sigma\approx9.4$ to $R_p/\sigma\approx10.2$ the packing fraction
changes from $\eta_p\approx0.54>\eta_0$ to $\eta_p\approx0.28<\eta_0$.  For our
setup, the lattice constant of the pattern is $20\sigma$. Hence,
the core of two neighboring wells mutually overlap if 
$R_p\ge R_t$, with $R_t/\sigma=10$. Therefore, if the well radius is
$R_p\gtrsim R_t-\sigma-2\delta\approx9\sigma$, it is possible to have bonds
between particles inside two different wells. See the snapshots shown in
Fig.~\ref{fig4}(b) state points 2 (non-percolated) and 3 (percolated). This is
the reason behind the abrupt decay of $\eta_p$. 

If the core of the wells overlaps, a percolation cluster can be formed by
particles all inside the core of the wells. As a result, the packing fraction
at the percolation transition decreases with respect to that on a clean
substrate. Also in this limit, it is possible to estimate the percolation
threshold. For that, we hypothesize that (i) the particles are all within the
core of the wells; (ii) the packing fraction inside the wells is $\eta_0$.
Then, the percolation transition is expected to occur at,
\begin{equation}
\eta_{\text{a,o}}(R_p)=A_o\eta_0,
\end{equation}
where $A_o$ is the area fraction of the substrate covered by overlapping
wells. A comparison between the predicted packing
fraction and the simulation results is shown in Fig.~\ref{fig4}(a).
Snapshots below and above the percolation threshold for the case of overlapping wells
are shown in Fig.~\ref{fig4}(b), state points $4$ and $5$, respectively. The
minimum threshold is obtained for $R_p/\sigma\approx 10.2$, for which
$\eta_p\approx 0.28$, what is $\sim 15\%$ lower than the one on a clean
substrate.
 
If the wells are very large, in comparison to the particle size, then the
entire substrate is covered by cores of the wells and the potential is uniform
across the substrate. In this limit, $A_o\rightarrow 1$, and we recover the percolation
threshold on a clean substrate, i.e.,
\begin{equation}
	\lim_{R_p\gg\sigma}\eta_p(R_p)=\eta_0.
\end{equation}

{\bf Repulsive patches.} The percolation phase diagram for a square lattice of
repulsive patches is shown in the left-hand side of Fig.~\ref{fig4}(a). The wells and
the interstitials interchange their role with respect to the case of
attractive wells. The wells are repulsive and hence the colloidal particles
tend to occupy first the interstitials until the density there is very high.
The limit of small repulsive wells is similar to the limit of very large
attractive wells. In both cases, we find $\eta_p<\eta_0$. The snapshots $6$
and $7$, in Fig.~\ref{fig4}(b), correspond to a non-percolated and a percolated
state, respectively, in the regime of non-overlapping repulsive wells. To
predict the percolation threshold for non-overlapping repulsive wells, we
hypothesize that (i) the particles completely avoid the repulsive wells; (ii)
in the interstitials the packing fraction is the same as in a clean substrate.
Then,
\begin{equation}
	\eta_{\text{r,no}}=(1-A_{\text{no}})\eta_0.
\end{equation}
This simple geometrical model slightly overestimates the effect of the
substrate on the transition (see Fig.~\ref{fig4}(a)), but still provides a
qualitative description of the percolation line. Note that, the above
hypotheses are valid only in the limit of a very strong substrate-particle
interaction, in which the particles completely avoid the repulsive regions.

The packing fraction at the percolation threshold decreases monotonically with $R_p$
until $R_p/\sigma\approx9$. For wells of size $R_p/\sigma=9$, the percolation
transition occurs at $\eta_p\approx0.19$, that is $\sim45\%$ lower than on a
clean substrate.

In the limit of large repulsive wells, particles tend to accumulate in the
interstitials, which are now isolated. This limit corresponds to what is
observed for small attractive wells. Therefore, as discussed above, to form a
percolation cluster, it is necessary to form bridges between interstitials.
The larger the interstitials are (without overlapping), the higher the packing
fraction is at the percolation threshold. An estimation for the packing
fraction at the percolation threshold is given by,
\begin{equation}
\eta_{\text{r,o}}=(1-A_{o})\eta_1+A_{o}\eta_0.
\end{equation}
To obtain this expression, we assumed that, at the percolation threshold, the
density in the interstitials is $\eta_1$ and the density inside the repulsive
wells is $\eta_0$. As before, setting
$\eta_1=0.65$, we obtain a good qualitative and quantitative agreement with
the numerical data, see Fig.~\ref{fig4}(a).

\begin{figure*}
\includegraphics[width=0.8\textwidth]{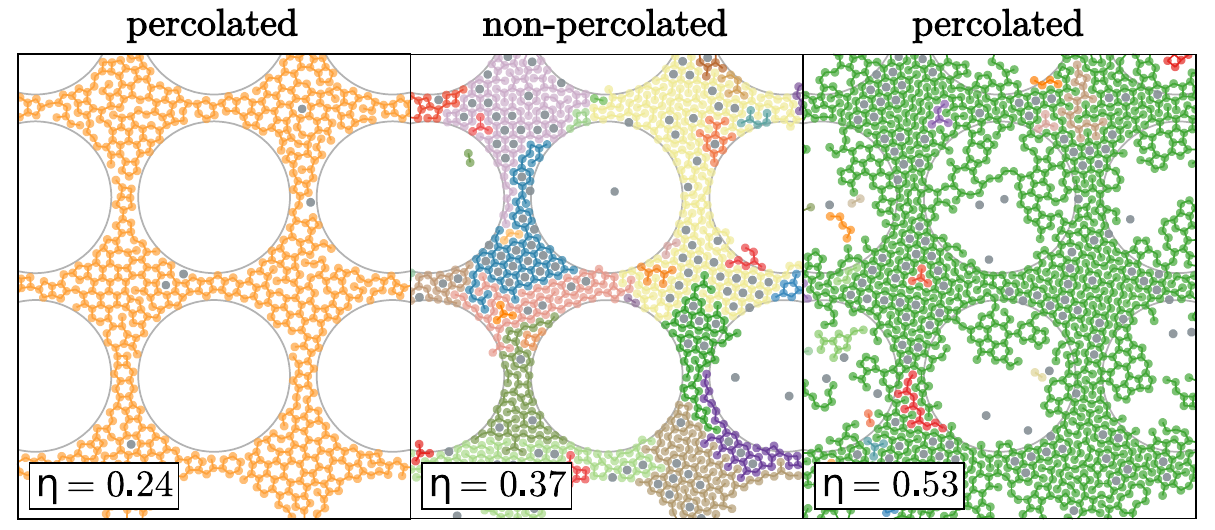}
\caption{Snapshots (partial region) for three different packing fractions, as
	indicated. The substrate is patterned with a square lattice of
	circular repulsive wells of radius $R_p/\sigma=8$. An arbitrary color
	has been assigned to each cluster. The system undergoes a reentrant
	percolation transition by increasing the packing
	fraction.~\label{fig5}}
\end{figure*}

We find a reentrant percolation transition for repulsive wells of size
$7\lesssim R_p/\sigma\lesssim9$, see Fig.~\ref{fig4}(a).  By increasing the
packing fraction (vertical direction in the diagram), we find: non-percolated,
percolated, non-percolated, percolated states. The snapshots $6$,$7$,$8$,
and $9$ in Fig.~\ref{fig4}(b), are representative of the reentrant
transition. The percolation at low packing fractions
(state points $6$ and $7$) corresponds to the one on a substrate of
non-overlapping repulsive wells, where particles accumulate in the
interstitials.  Surprisingly, by increasing the packing fraction, the system
undergoes a transition from a percolated to a non-percolated state (see state
points $7$ and $8$). This transition is driven by the formation of defects on
both the orientational and the spatial order of the particles.  To accommodate
an increase in the density of particles in the interstitials, the particles
organize in a compact manner, with local hexagonal order, as indicated by the
high value of $\tilde q_6$ (state points $8$ and $9$). The particles form also
a fully bonded honeycomb-like structure with isolated particles in the middle
of each unit cell. As shown in Figs.~\ref{fig4}(b)~and~\ref{fig5}, the
interstitials are interconnected through narrow channels, with a width of a
few particle diameters. Through these channels, the formation of a percolation
cluster might be suppressed due to the emergence of bond and structural defects.
The formation of structural defects is consistent with the observed
drop of the order parameter $\tilde q_6$, see state points $8$ and
$9$. The bond defects are visible in the snapshots shown in Fig.~\ref{fig5}, where
different clusters have different colors. For even larger packing fractions,
the percolation is recovered via the formation of bridges inside the repulsive
wells, see state point $9$, in Fig.~\ref{fig4}(b).

\begin{figure}
\includegraphics[width=1.0\columnwidth]{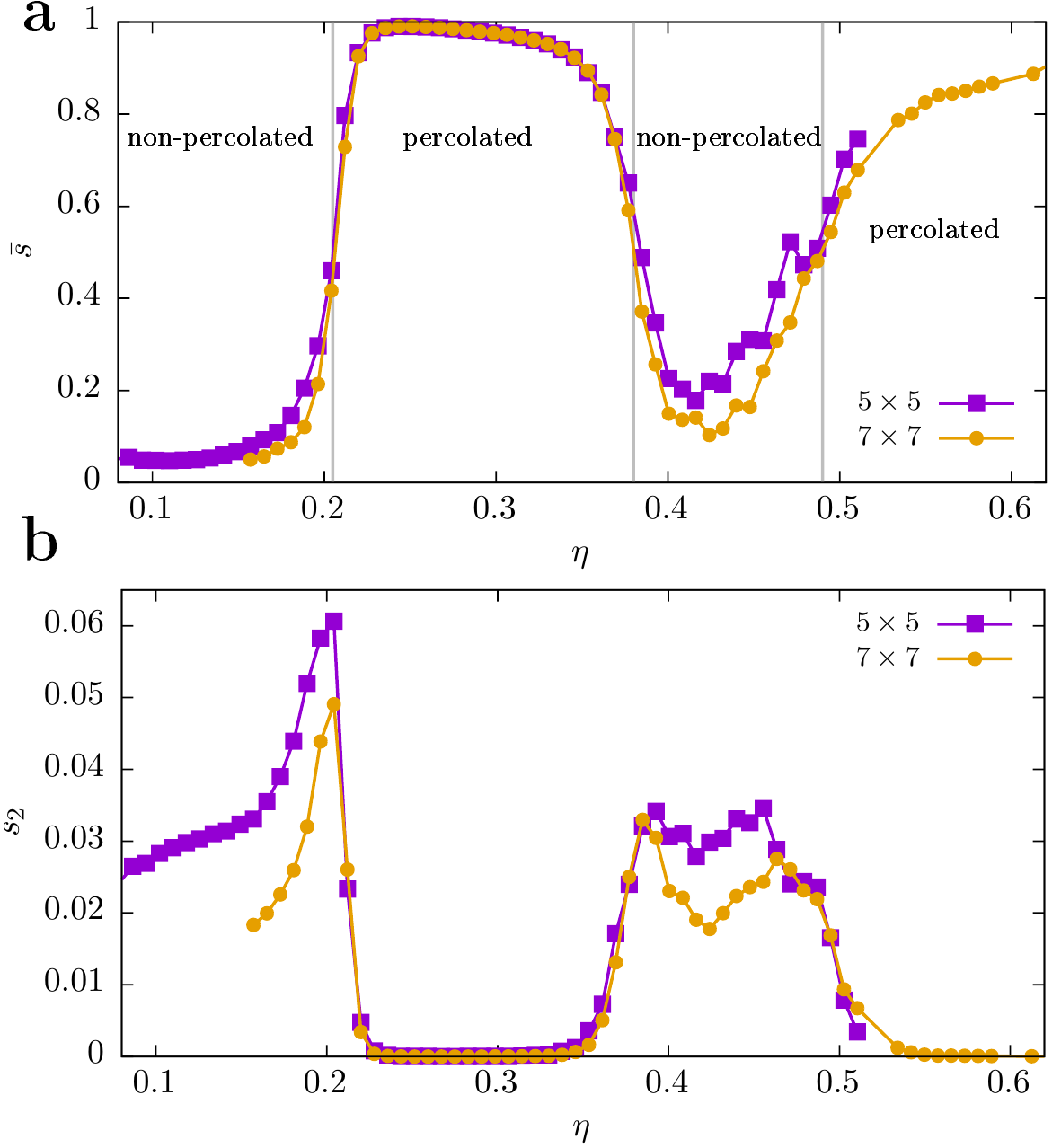}
	\caption{Fraction of particles in the largest cluster (a) and second
	moment of the cluster-size distribution (b) as a function of the
	packing fraction. The substrate consists of a square lattice of
	$n_p\times n_p$ repulsive wells, each of size $R_p/\sigma=8$. Two
	system sizes are shown: $n_p=5$ (violet squares) and $n_p=7$ (orange
	circles). The gray vertical lines in (a) indicate the packing fraction
	at which $\bar s=0.5$, what roughly coincides with the maximum in
	$s_2$.~\label{fig6}}
\end{figure}

In Fig.~\ref{fig6}, we show the fraction of particles in the largest cluster
(a) and the second moment of the cluster-size distribution (b) in a substrate
with repulsive wells of $R_p/\sigma=8$. Both quantities are consistent with a
reentrant transition. We have considered different system sizes and no strong
finite-size effects are observed (see figure). 

\subsection{Percolation on a patterned substrate: hexagonal lattice}
\begin{figure*}
\includegraphics[width=0.8\textwidth]{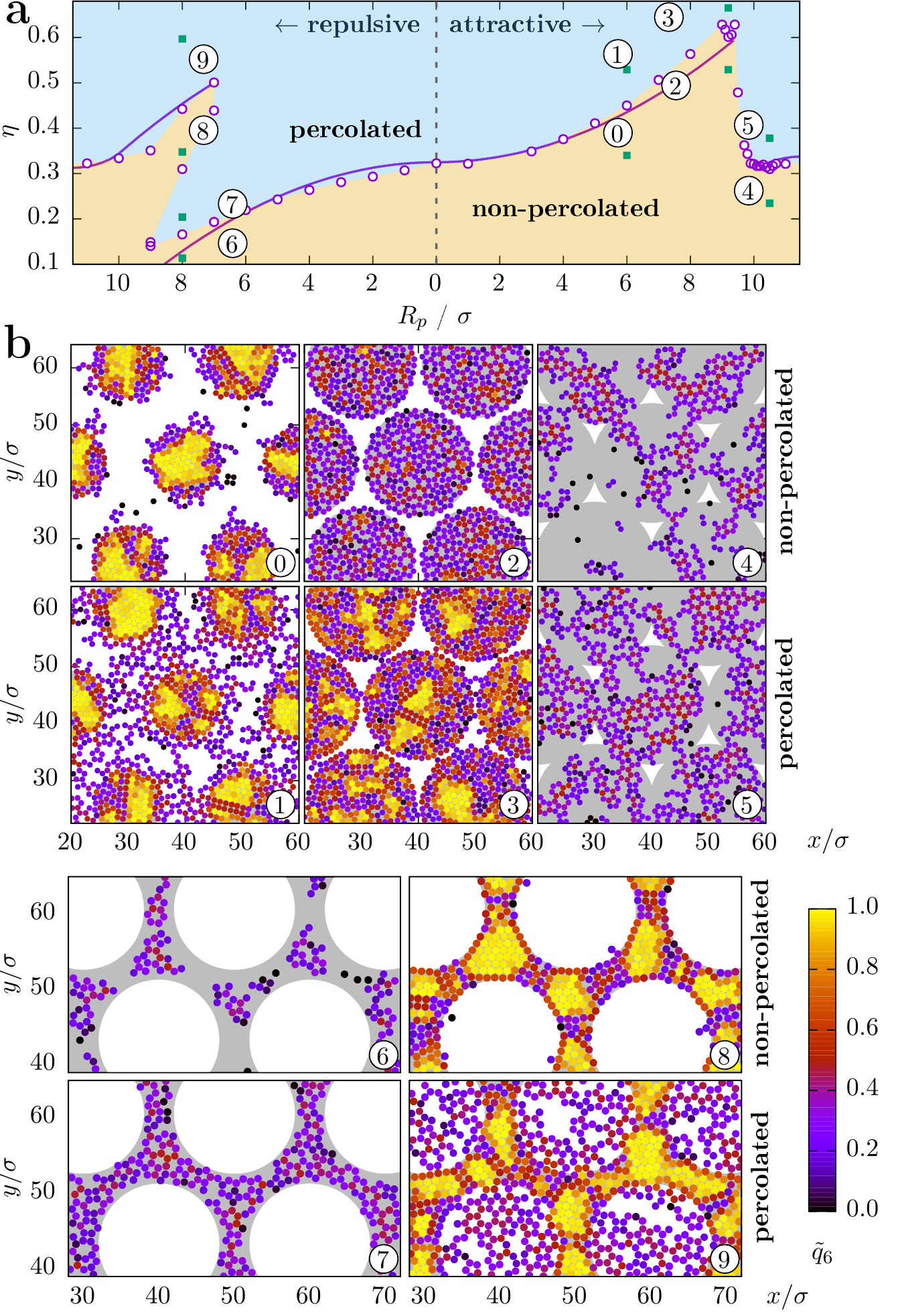}
\caption{Hexagonal lattice. (a) Percolation phase diagram in the plane of
	radius of the potential well core $R_p$ and packing fraction $\eta$.
	The light brown (light blue) area is the non-percolated (percolated)
	region, as indicated. On the left (right) of the vertical dashed line
	we show the results for repulsive (attractive) wells. The wells on the
	substrate form a hexagonal lattice. The numbered green squares are
	selected state points for which a snapshot is shown in panel (b). The
	empty circles are the percolation values estimated numerically using
	Monte Carlo simulations. (b)
	Selected snapshots of the central region of the substrate.  The
	packing fractions and radii of the well of each configuration is
	indicated in panel (a) (green circles). The top (bottom) group of
	snapshots correspond to attractive (repulsive) wells. The grey areas
	indicate the energetically favorable regions for the colloidal cores.
	The colloids are colored according to the value of their order
	parameter $\tilde q_6$.\label{fig7}}
\end{figure*}

We now consider a pattern consisting of circular wells arranged in a hexagonal
lattice. The percolation phase diagram together with snapshots are shown in
Fig.~\ref{fig7}(a)~and~(b), respectively.

The percolation phase diagram is qualitatively the same as in the case of a square
lattice. The simple geometrical models also predicts quantitatively the packing
fraction at the percolation transition. The reentrant percolation transition
for repulsive wells is also observed. As in the previous substrate pattern,
the formation of defects drives a transition to a non-percolated state by
increasing the density from a percolated state. 

\section{Summary and conclusions}
The presence of a patterned substrate substantially affects the percolation
properties of a system of functionalized colloidal particles. We have studied
substrate patterns consisting of a sequence of either attractive or repulsive
circular wells.  For increasing packing fractions, the colloidal particles
accumulate first in the regions of lower potential energy, i.e., inside the
wells if they are attractive or in between if they are repulsive. The regions
of high potential energy are only occupied for large packing fractions. When
the regions of lower potential energy percolate (e.g. overlapping attractive wells),
the percolation transition for the colloidal particles occurs for lower packing
fractions than in the case of a clean (non-patterned) substrate. On the contrary,
if the low potential energy regions are fragmented,
the percolation threshold is indeed larger than in a clean substrate.
We have shown that the percolation threshold can be estimated by a simple model, based on pure
geometric arguments. 

In the case of repulsive wells, we find a reentrant percolation transition
related to the formation of structural and bond defects, what suppresses
percolation for intermediate values of packing fraction. The percolation is
only recovered for very high packing fractions. Previous studies of functionalized
colloidal particles have reported also reentrant phenomena driven by, e.g.,
gravitational fields~\cite{PhysRevE.93.030601,0953-8984-29-6-064006}, competition
between different energy
scales~\cite{PhysRevLett.106.085703,doi:10.1063/1.4819058,doi:10.1063/1.4849115}, and patches
activated by temperature~\cite{0953-8984-28-24-244008}.

For non-overlapping wells, the cluster-size distribution is characterized by
well-defined peaks at certain cluster sizes. This suggests that the percolation
transition might be no longer continuous~\cite{PhysRevLett.105.035701}. Future works
might consider studying the nature of the transition in detail for different
patterns, temperatures, and number of patches per particle.

For simplicity, we considered a two-dimensional system but we expect the
results to be valid in the sub-monolayer regime, provided that the vertical
position of the particles relative to the substrate does not vary
significantly. There are several experimental techniques available to obtain
patterned substrates, such as, e.g., microcontact
printing~\cite{doi:10.1021/ar00053a003}, chemically patterned substrates with
anionic and cationic regions~\cite{PhysRevLett.84.2997}, soft
lithography~\cite{doi:10.1021/la030380c}, and optical substrates using arrays
of optical
tweezers~\cite{doi:10.1063/1.1148883,PhysRevLett.89.128301,doi:10.1063/1.1488690,doi:10.1021/la990610g}.
Also various approaches have been developed to synthesize functionalized
colloids~\cite{in2,0953-8984-25-19-193101,gong2017patchy,BIANCHI20178,B716393B}.

We have considered a monocomponent system. New phenomenology is expected in
the case of binary and ternary mixtures of functionalized colloids for which
several types of percolated states are possible. Examples are
bigels~\cite{paper14,Varrato20112012} and trigels~\cite{paper32} in which
interpenetrated networks made of single species percolate the system. 

Finally, we have focused on equilibrium properties. Some of the phenomenology
described here might be dynamically inaccessible due to the formation of
kinetically trapped structures~\cite{Dias2016,Araujo2017,Dias2017,Dias2018}.
Analyzing the dynamics on patterned substrates is therefore a fundamental
question of practical interest that should be considered in future works.

\acknowledgments This work was partially funded by the Portuguese-German
FCT/DAAD project ``Self-organization of colloidal particles on patterned
substrate'' (DAAD project-id: 57339919). NA also thanks financial support from
the Portuguese Foundation for Science and Technology (FCT) under the contract
no. UID/FIS/00618/2013.

\end{document}